\documentclass[pre,aps,showpacs,twocolumn,preprintnumbers,amsmath,amssymb,superscriptaddress, floatfix]{revtex4}

\usepackage [caption=false] {subfig}
\usepackage{graphicx}
\usepackage{epstopdf}
\usepackage{epsfig}
\usepackage{graphics}
\usepackage{dcolumn}
\usepackage{bm}
\usepackage{algpseudocode}
\usepackage{enumitem}
\usepackage{float}
\usepackage{xcolor}

\newcommand{\I}{\mathcal{I}}
\newtheorem{theorem}{Theorem}
\newtheorem{example}[theorem]{Example}
\renewcommand{\bar}{\overline}
\DeclareMathOperator*{\argmax}{argmax}
\DeclareMathOperator*{\argmin}{argmin}

\begin{document}


\title{Artificial intelligence applied to bailout decisions in financial systemic risk management}

\author{Daniele Petrone}
\author{Neofytos Rodosthenous}  
\author{Vito Latora} 

\affiliation{School of Mathematical Sciences, Queen Mary University of London. Mile End Road, London E1 4NS, UK}
\email{v.latora@qmul.ac.uk}

\date{\today}

\begin{abstract}
We describe the bailout of banks by governments as a Markov Decision Process (MDP) where the actions are equity investments
. The underlying dynamics is derived from the network of financial
institutions linked by mutual exposures, and the negative rewards are associated to the banks' default.
Each node represents a bank and is associated to a probability of default per unit time (PD) that depends on its capital and is increased by the default of neighbouring nodes. Governments can control the systemic risk of the network by providing additional capital to the banks, lowering their PD at the expense of an increased exposure in case of their failure.   
Considering the network of European global systemically important institutions, we find the optimal investment policy that solves the MDP, providing direct indications to governments and regulators on the best way of action to limit the effects of financial crises.   
\end{abstract}

\maketitle

\section{Introduction}

In times of crisis, as during the recession of 2008 or the economic disruption triggered by the COVID-19 pandemic, the governments face difficult decisions regarding bailing-out strategically important companies. In particular, large banks are critical for the stability of the financial system and are closely monitored by central banks and governments. As an example, to rescue Royal Bank of Scotland (RBS) in 2008-2009, the UK government became the majority shareholder of the bank, purchasing shares for a total 45.5 billion pounds 
\cite{RBS_ShareOwnership_2020}. The government achieved its objectives to stabilise the financial system, and no depositor in UK banks lost money. However, the cost for taxpayers has been estimated by the Office for Budget Responsibility (OBR) to be in the region of 27 billion pounds as of March 2018 \cite{BOR_EFO_2018}. The price of RBS shares plummeted after the purchase and the government has since sold part of its investment at a loss.
Was the government intervention value for money? The National Audit Office (NAO) is the UK's public spending watchdog and in December 2009 released the report ``Maintaining financial stability across the United Kingdom’s banking system" \cite{NAO_Managing_2009} where they analysed the government support for the banking sector and the conclusion was that: ``If the support measures had not been put in place, the scale of the economic and social costs if one or more major UK banks had collapsed is difficult to envision. The support provided to the banks was therefore justified, but the final cost to the taxpayer of the support will not be known for a number of years". 
NAO did not produce an estimate of the impact in case of inaction of the government.
In this paper, we propose a mathematical framework that allows a quantitative comparison between investment decisions by the government.
Our framework is based on the three following building blocks: 
$(a)$ a dynamical network model of the financial system with a contagion mechanism between financial institutions; 
$(b)$ a set of allowed government interventions to control the network; and 
$(c)$ a quantitative way to assess the government actions at each time step. 
A network model \cite{Caccioli_Network_2018}\cite{boccaletti_complex_2006}\cite{lehar_measuring_2005}\cite{furfine_interbank_2003} is essential, as the main concern is not the direct cost of a default but the systemic risk that it entails. 
Systemic risk can be defined as the risk that large part of the financial system is disrupted and as such it requires connections between financial institutions that can transfer the distress along the network \cite{gai_contagion_2010}\cite{haldane_managing_2014}\cite{battiston_debtrank_2012}. The contagion mechanism that we use is the impact 
that a bank default has on other banks \cite{petroneLatora_dynamic_2018}. The impact can be due to direct losses in bilateral credit exposures \cite{upper_estimating_2004}\cite{upper_simulation_2011} (for example if they had lent money to the defaulting bank), or indirect losses due to fire selling of assets by the defaulting bank \cite{wagalath_running_2011}, that would lower the market value of similar assets in the balance sheet of the other financial institutions. The impact would lower the capital buffer of the affected banks, 
weakening the network and its ability to withstand future shocks. In particular, the probability of default per unit time (PD) of the nodes (banks or financial institutions) would increase, hence increasing the expected loss in the network \cite{petroneLatora_dynamic_2018}. 
One main novelty of our model is that we allow for the network to 
be controlled by a government investment in the capital of the banks. 
Such an investment would, conversely, decrease 
the PD of the banks that receive the additional capital, but also increase the loss for the government in case of default.
In our framework, the connection between the change in PD and the variation in the amount of capital is provided by the Merton model of credit risk \cite{merton_pricing_1974}
. To follow the evolution in time of the network, we simulate the default process given the PD of the nodes and their tendency of defaulting during the same time step. 
Finally, we use artificial intelligence techniques \cite{sutton_reinforcement_2018}\cite{ohalloran_artificial_2019}\cite{kou_machine_2019} to assess the optimality of government decisions (no investment vs different amounts of investment), recasting the system as a Markov Decision Process (MDP) \cite{bellman_markovian_1957} where the actions (controls) are government investments at each time step.

The paper is structured as follows. In section~\ref{ssec:network} we describe the network of financial institutions, its dynamics and contagion mechanism, and then in section~\ref{subSec:MDPdefinition} we introduce a Markov Decision Process based on the 
network, in order to model government interventions on bailed-out banks. 
We continue in section~\ref{subSec:SolvingMDP} with presenting our strategy to solve the MDP, by finding the optimal government investment decision for each state of the network and time. Section~\ref{sec:Results} contains our results, obtained by applying our model to a homogeneous network organised as a
Krackhardt kite graph (see Fig. \ref{fig:KKpgraph}) and to the network of the European Global Systemically Important Institutions. 
We have found that a pre-existing investment in a distressed node makes it convenient for the government to intervene again to try to save the invested capital (creating moral hazard as the node could act haphazardly relying on the implicit government guarantee). 
Moreover, by changing the parameter $\alpha$, that accounts for the taxpayers' loss in case a bank defaults, we have observed that there is a `critical' value that separates networks for which the inaction of the government is the best option from networks where an investment of the government would be the optimal decision as it would lower the overall expected loss of the system. 
Finally, we provide our conclusions in section~\ref{sec:ConcludingRemarks}.

\section{Our framework}
\subsection{Network of financial institutions}
\label{ssec:network}

We consider a network $G$ with  a set $\I = \{1,...,N\}$ of nodes representing financial institutions. Each node $i \in \I$ is characterised at time $t$ by a probability of default $PD_i(t) \in (0,1]$ per time interval $\Delta t$, a total asset $W_i(t)$ and an equity $E_i(t)$ (such that $E_i(t) \leq W_i(t)$), 
that is the capital used by node $i$ as a buffer to withstand financial losses 
. 
The edges $w_{ij}$ of the network represent the exposure of node $i$ to the default of node $j$ 
for all 
$i\not= j \in \I$.
To take into account government interventions aimed at limiting the overall losses, we use an adaptation of the `PD Model' described in \cite{petroneLatora_dynamic_2018} by extending it to allow the possibility for the nodes  (banks) to incur positive shocks, via investments in the nodes, rather than just negative shocks due to the default of other nodes.
The focus has also changed from the one in \cite{petroneLatora_dynamic_2018}, as we are now exclusively interested in the losses incurred by the taxpayers, disregarding the losses sustained by private investors. 
In the following, we will measure the time in discrete time steps that are 
multiples of $\Delta t$, i.e. $t + 1$ 
is equivalent to $t + \Delta t$.
 
We define the total impact $I_i(t)$ on node $i$ at time $t$, due to the default of other nodes $j\in \I \setminus \{i\}$ 
in the network as
\begin{align}
	I_i(t) &:= \sum_{j\in \I \setminus \{i\}} 
	w_{ij}(t) \delta_j(t) 
, \quad \text{for all } i\in \I,
	\label{eq:Impact} 	
\end{align}
 where $\delta_j(t)=1$ if and only if node $j$ defaults at time $t$ and $\delta_j(t)=0$ otherwise.
 The impact $I_i(t)$ represents a loss for the total asset $W_i$, which in turn decreases also the equity $E_i$ of node $i$, hence reducing their value at time $t+1$. 
 This can be seen from the accounting equation 
for each node $i$, namely
  \begin{align}
  	 W_i(t) &= E_i(t) + B_i(t) \,,
  	 \label{eq:AccountingEquation}	 	
  \end{align}
 which states that the total asset $W_i$ is always equal at all times to the equity $E_i$ plus the total liability $B_i$. Note that $B_i$ is not affected by the losses as it is comprised of loans from other banks, deposits, etc., that are due in full unless the bank $i$ defaults. 
 Hence, we have
   \begin{align}
   	\Delta W_i(t)  &=  \Delta E_i(t) \,,
   	\label{eq:DeltaAccountingEquation}	 	
   \end{align}     
where we define $\Delta X_i(t):= X_i(t+1) - X_i(t)$. 
We can therefore write 
\begin{align}
 	\label{eq:DeltaW}	
 	W_i(t+1) - W_i(t)  &= - I_i(t) + \Delta J_i(t), \\
 	\label{eq:DeltaE}
	E_i(t+1) - E_i(t)  &= - I_i(t) + \Delta J_i(t),
\end{align}
where $\Delta J_i(t)$ denotes the potential increase in the current investment $J_i(t)$ of the government in node $i$ at time $t$. 
On the other hand, the probability of default $PD_i(t)$ of node $i$ is increased by the impact $I_i(t)$ at time $t$, since part of the capital buffer (equity $E_i$) is lost. 
In order to model the effect of the impact $I_i(t)$ on $PD_i(t)$, we use the  Merton model for credit risk \cite{merton_pricing_1974} to calculate 
the `implied probability of default' $PDM$ as a function 
of the parameters of each node:
\small
\begin{align} \label{eq:PDMLongVersion}
PDM(W,E,\mu, \sigma) := 1- \Phi\Big( \Big(log\frac{W}{W-E} + \mu - \frac{\sigma^2}2 \Big)\big/ \sigma \Big)
\end{align}
\normalsize
where the term $W - E$ represents the total liability $B$ of each bank, $\Phi$ is the univariate standard Gaussian distribution, $\mu$ is the drift and $\sigma$ is the volatility of the geometric Brownian motion associated to the total asset $W$ in the Merton model. 
We then use \eqref{eq:PDMLongVersion} to obtain  
\small
\begin{align}
PD_i(t) := max\{PDM(W_i(t),E_i(t),\mu_i,\sigma_i),PDM^{floor}_i \}
	\label{eq:PDM}
\end{align}
\normalsize
where we introduced the fixed number ``$PDM^{floor}_i$" representing the lower bound of the $PD_i$ that is used to exclude unreasonably low probabilities of default.
For example, it is a standard assumption for the $PD_i$ of a bank $i$ to be greater or equal to the probability of default of the country where it is based. In this context, the latter is the probability of a country defaulting on its debt.

Now, if node $i$ loses an amount of capital $I_i(t)$ at some time $t$ 
greater or equal to its buffer $E_i(t)$, the total asset $W_i(t)$ becomes less than its liability $B_i(t)$ and it is convenient for the shareholders to exercise their option to default. 
In practice, when this occurs, we set $PD_i(t+1) = 1$ and node $i$ will default at time $t+1$. 
Moreover, recall that node $i$ may also default at any time $t$ with probability $PD_i(t)$ due to its own individual characteristics given by \eqref{eq:PDM}; see also the default mechanism in \eqref{eq:defaultCondition} below.

Now, when node $i$ defaults, we denote by $LGD_i$ the ``Loss Given Default" of node $i$, which is a fixed number representing the percentage of the investments $J_i$ on node $i$ by the government, that cannot be recovered after a default.
In case of default of node $i$, we further assume that in addition to the aforementioned loss of investments, the taxpayers' loss $L_i$ is also comprised of a fixed percentage $\alpha_i$ (for convenience) of the total asset $W_i$ of the node. That is, the taxpayers' overall loss $L_i$ is given by 
\begin{align} 
   L_i :=  \alpha_i \; W_i + J_i \; LGD_i \,. 
  \label{eq:GovLoss}
\end{align}

To complete our framework we need to specify the probability of more than one default happening during the same time step, given the $PD_i$ of each node $i$ obtained by \eqref{eq:PDM}. For example, if the nodes were independent the probability of nodes $i$ and $j$ defaulting at the same time step, denoted by $PD_{[ij]}$, would be the product of the individual probabilities $PD_i$ and $PD_j$. 
In this paper, we allow nodes to depend on each other and use a Gaussian latent variable model \cite{okane_gaussian_2008} 
to calculate the probabilities of simultaneous defaults of two or more nodes. To be more precise, the probability of a finite subset of nodes $\{i,j,k,...\} \subseteq \I$ in the network G defaulting at the same time, is given by the following integral
\begin{align} 
	PD_{[i,j,k,...]} &:= \int_D \Phi_N' ({\bf u} ;\Sigma) 
	\; d{\bf u} \,,
	\label{eq:multipleDef}
\end{align}
where $\Phi_N'$ is the standardised multivariate 
Gaussian density function with zero mean and 
a symmetric correlation matrix $\Sigma \in [-1,1]^{N \times N}$ given by
\begin{align} 
 \Phi_N' ({\bf u};\Sigma) &:= \frac{\exp\{ -\frac12 {\bf u}^T \Sigma^{-1} {\bf u} \}}{\sqrt{(2 \pi)^n \, |\Sigma|}}
	\label{eq:multivariateDensity}
\end{align}
and $|\Sigma|$ is the determinant of 
$\Sigma$.
We further note that the integration domain $D$ in \eqref{eq:multipleDef} is the Cartesian product of the intervals $[-\infty, \Phi_1^{-1}(PD_i)]$ for each node $i$ that belongs to the set of defaulting nodes, and the intervals $[-\infty, \infty]$ for the remaining nodes, where $\Phi_1$ is the univariate standard Gaussian distribution.  

In the sequel, this model will also be used to simulate the default mechanism. To be more precise, by sampling values $x_1, ..., x_N$ of the random vector $X = (X_1, X_2, ..., X_N)^T$ with the multivariate Gaussian distribution mentioned above, at each time step $t$, we will assume that node $i$ defaults according to the rule:
\begin{align} 
	x_i  \; <  \Phi_1^{-1}(PD_i(t)) \iff   \delta_i(t) = 1 \,.   
	\label{eq:defaultCondition}
\end{align}   

\subsection{Formulation of the banks bailout problem as a Markov Decision Process}
\label{subSec:MDPdefinition}
We describe the government decisions of bailing out banks as a Markov Decision Process (MDP) driven by the network framework described above. We assume that the government estimated that the crisis will likely be over at time M, and in any case it will be able to sell the shares of the rescued banks to the private sector for a price that is similar to the purchasing price.
We define the 4-tuple $(S, A_s, P_a, R_a)$ of the set $S$ of all the states, 
set $A_s$ of all
actions available from state $s\in S$, transition probabilities $P_{a}(s,s') = P(s_{t+1}= s'| s_t = s, a_t = a )$ between state s at any time $t$ and state $s'$ at time $t+1$ having taken action $a\in A_s$ at time $t$, and rewards (negative losses in our model) $R_{a}(s,s')$ received after taking 
action $a$ at any time $t$ while being at state $s$ and landing in state $s'$ at time $t+1$,  where $s,s'\in S$. 
Furthermore, a constant discount factor $\gamma$ is defined with $0 \leq \gamma < 1$, so that rewards obtained sooner are more relevant in the calculation of the cumulative reward CR over M steps. 
The latter is therefore defined by
\begin{align}
	CR := \sum_{t=0}^{M-1} \gamma^t R_{a_t} (s_t, s'_{t+1}) .
	\label {eq:CumulativeReturn}
\end{align}	
In the remaining of this section, we expand on the 4-tuple $(S, A_s, P_a, R_a)$ that defines our MDP.
\vspace{3pt}

{\em MDP states.} 
The states $s_t \in S$, at each time $t$, are defined by three main pilars: $(a)$ all the parameters of the network $G$ ($W_i(t)$, $E_i(t)$, $PD_i(t)$, $LGD_i$, $\alpha_i$, $\mu_i$, $J_i$, $\sigma_i$, $w_{ij}$, $\Sigma_{ij}$, for $i,j \in \{1,...,N\}$, where $w_{ii} = 0$), $(b)$ an indexed set $\I_{def}(t) \subseteq \I$ containing all defaulted nodes 
prior to time $t$ and $(c)$ the time to maturity $M-t$. 
\vspace{3pt}

{\em MDP actions.} 
The MDP actions in our model are injections of capital $a_t \rightarrow \Delta J^a(t) = (\Delta J_1^a(t), \Delta J_2^a(t),..., \Delta J_N^a(t))$ by the government to the nodes $(1,2,...,N)$. These additional resources on one hand, make the nodes more resilient, hence diminishing their probability of default via \eqref{eq:DeltaW}--\eqref{eq:PDM}, but on the other hand they will be at risk in case of default since they increase each 
$J_i$ in \eqref{eq:GovLoss}.  
These actions are the control variables of the government when trying to minimise the losses of the network (i.e. maximise the expected CR in \eqref{eq:CumulativeReturn}, see Section \ref{subSec:SolvingMDP} for more details).

We further assume that these  government investments (relative to action $a_t$) decided at time $t$ are implemented immediately, so that the probability of default ${PD}_i(t)$ given by  \eqref{eq:PDM}, the default mechanism in \eqref{eq:defaultCondition} and subsequently the impacts $I_i(t)$, for each node $i\in\I$, are implemented using the updated (increased) capital $\bar{E}_i(t) = E_i(t) + \Delta J_i^a(t)$, total asset $\bar{W}_i(t) = W_i(t) + \Delta J_i^a(t)$ and government investment $\bar{J}_i(t) = J_i(t) + \Delta J_i^a(t)$. 
\vspace{3pt}

{\em MDP transition probabilities.}
Within our framework, a node that has defaulted does not contribute to future losses and cannot become active again, i.e. the  cardinality  of  the  set of defaulted nodes $|\I_{def}(t)|$ is a non-decreasing function of time $t$. 
Hence the transition probability $P_{a}(s,s')$ from state $s$ to $s'$ will be non-zero only for states $s'$ that: 
$(a)$ have the same number or more defaulted nodes than state $s$;
$(b)$ are ``reachable", in the sense that their $PD_i(t+1)$, $W_i(t+1)$ and $E_i(t+1)$, 
for $i \in \I\setminus\I_{def}(t+1)$ (the set of remaining active nodes in $s'$)
take values that are coherent with equations~\eqref{eq:DeltaW}--\eqref{eq:PDM} after calculating the impacts $I_i(t)$ from the nodes $i\in\I_{def}(t+1)\setminus\I_{def}(t)$. 
In order to illustrate the above we consider the following example.
\begin{example}
\normalfont
Let us consider a network with three nodes $\I=\{1,2,3\}$ and $w_{ij} = 1 \; \forall\, i\not=j\in\I$, at a time $t$, such that node $3 \in \I_{def}(t)$ has already defaulted, while the remaining nodes have $W_i(t) = 100$, $E_i(t) = 3$ and $PD_i(t) = 0.001$ for $i\in\I\setminus\I_{def}(t)=\{1,2\}$. 
In case the government does not intervene, the states $s'$ that can be reached are the ones where: 
$(i)$ all the nodes default at time $t$, i.e. $\I_{def}(t+1)=\I$; 
$(ii)$ nodes $1$ and $2$ are still active and $W_i(t+1)$, $E_i(t+1)$ and $PD_i(t+1)$ for $i\in\{1,2\}$ are the same as for state $s$; 
$(iii)$ node $1$ defaults at time $t$ while node $2$ remains active, i.e. $\I_{def}(t+1)=\{1,3\}$, $W_2(t+1) = 99$ and $E_2(t+1) = 2$ (since the impact $I_2(t)= w_{21}= 1$) and $PD_2(t+1)$ needs to take the value calculated via \eqref{eq:PDM} using the $W_2(t+1)$ and $E_2(t+1)$ inputs; 
and $(iv)$ node $2$ defaults at time $t$ but node $1$ remains active, which is analogous to $(iii)$ by swapping indices $1$ and $2$.
Now, if the government decides to invest, i.e. $a \rightarrow (\Delta J^a_1(t), \Delta J^a_2(t))$ on nodes $1$ and $2$, respectively, at time $t$, we need to update the capitals $\bar{E}_i(t) = E_i(t) + \Delta J^a_i(t)$ and total assets $\bar{W}_i(t) = W_i(t) +  \Delta J^a_i(t)$ for $i\in\{1,2\}$ according to the government intervention and then use the updated $\bar{E}_i(t), \bar{W}_i(t)$ to perform the same analysis as above to identify the reachable states. 
\end{example}
For states $s'_{t+1}$ with a non-zero transition probability $P_{a_t}(s_t,s'_{t+1})$, we can calculate the latter via the Gaussian latent variable model, thus they will depend exclusively on the parameters $PD_i$ and $\Sigma_{ij}$ with $i,j \in  \I\setminus\I_{def}(t)$. 
To be more precise, we first create an intermediate state $\bar{s}$ by applying the government investments relative to action $a_t$ to state $s_t$; hence each node $i$ of $\bar{s}$ will have an increased capital $\bar{E}_i(t) = E_i(t) + \Delta J^a_i(t)$, an increased total asset $\bar{W}_i(t) = W_i(t) + \Delta J^a_i(t)$, an increased government investment $\bar{J}_i(t) = J_i(t) + \Delta J^a_i(t)$ and a probability of default given by  \eqref{eq:PDM} with inputs $\bar{W}_i(t)$ and $\bar{E}_i(t)$. 
Using the intermediate state $\bar{s}$ with updated $\bar{E}_i(t), \bar{W}_i(t), \bar{J}_i(t)$ and updated ${PD}_i(t)$, we calculate the transition probability 
via (see also \eqref{eq:multipleDef}) the following integral 
\begin{align}
	P_{a_t}(s_t, s'_{t+1})	:= \int_D \Phi_{|\I\setminus\I_{def}(t)|}' ({\bf u}; \Sigma_{sub}) d{\bf u} ,
	\label{eq:transProb}
\end{align}
where $\Phi'$ is the 
density given by \eqref{eq:multivariateDensity} with dimension 
equal to the cardinality of the set of surviving nodes $|\I\setminus\I_{def}(t)| \leq N$.
Moreover, the integration domain $D$ in \eqref{eq:transProb} is the Cartesian product of the intervals $[-\infty, \Phi_1^{-1}({PD}_i)]$ for the additional defaulted nodes $i \in \I_{def}(t+1)\setminus\I_{def}(t)$ 
and the intervals $[\Phi_1^{-1}({PD}_i), \infty]$ for all the remaining active nodes $i \in \I\setminus\I_{def}(t+1)$ at state $s'_{t+1}$ 
-- upon recalling the default mechanism in \eqref{eq:defaultCondition}. 
The $\Sigma_{sub}$ is the sub-matrix of the original correlation matrix $\Sigma$ after removing the rows and the columns corresponding to defaulted nodes $i \in \I_{def}(t)$ at state $s_t$.  
\vspace{3pt}

{\em MDP rewards.}
In our model the ``rewards" take non-positive values, since their overall maximisation has to translate for our MDP into the minimisation of the overall taxpayers' losses $L_i(t)$ given by \eqref{eq:GovLoss}, for all nodes $i\in\I\setminus\I_{def}(t)$ at each time $t$. Namely, 
\small
\begin{align}
 R_{a_t}(s_t,s'_{t+1})  :=  - \hspace{-7pt} \sum_{i\in\I\setminus\I_{def}(t)} \hspace{-7pt} (\alpha_i \; \bar{W}_i(t) + \bar{J}_i(t) \; LGD_i) \; \delta_i(t) ,
 \label{eq:MDPRewardDefinition}
\end{align}
\normalsize 
where only the nodes defaulting at time $t$ with $\delta_i(t) = 1$ contribute to the sum of losses, hence the ``reward" is $0$ if there are no additional defaults at time $t$. 
The expected ``reward" depends on the action taken, as it influences the total asset $\bar{W}_i$ and investment $\bar{J}_i$ corresponding to the intermediate state $\bar{s}$ described previously, as well as the $PD_i$, i.e. the probability of having $\delta_i(t) = 1$.
 
\subsection{Solving the Markov Decision Process}
\label{subSec:SolvingMDP}
Solving the MDP means to find the optimal action for each possible state $s_t$. In our context, we expect our model to indicate if the government should intervene and if so, which amount it should invest for a given configuration of the financial system network. To find a solution and describe it mathematically, we need to define a few concepts as described below.

\vspace{3pt}
{\em Optimal policy.}
The optimal policy $\pi_* (s_t) \rightarrow a^*_t$, is a function that returns the optimal action $a^* \in A_{s}$ for each state $s$ at time $t$. The optimal action is the one that obtains the maximum expected cumulative reward as defined in \eqref{eq:CumulativeReturn}.

\vspace{3pt}
{\em Optimal value function.}
The optimal value function $V_*(s_t)$ is 
defined by 
\begin{align}
	V_*(s_t) := E_{\pi_*}[CR\,|\,s_t].
	\label{eq:MDPVstarDefinition}
\end{align} 
This is the expected cumulative reward starting from state $s_t$ and following the optimal policy $\pi_*$ for any of the successive time steps till the end of the episode (recall that a full episode consists of M time steps). One way to obtain this expected cumulative reward is to run the MDP starting at $s_t$ multiple times and average the results.
Given the definition of $\pi_*$, $V_*(s_t)$ represents the maximum expected cumulative reward that can be obtained starting from $s_t$. 

\vspace{3pt}
{\em Optimal action value function.}
The optimal action value function $Q_*(s_t, a_t)$ is the expected cumulative reward we obtain if we first take action $a_t$ at state $s_t$ and then follow the optimal policy $\pi_*$ for any of the successive steps from $t+1$ until the end of the episode $M$. It is defined by
\begin{align} \label{Qas}
Q_*(s_t, a_t) &:= E_{\pi_*}[CR\,|\, s_t, a_t]. 
\end{align}
Similarly to the previous paragraph, this represents the maximum expected cumulative reward that can be obtained starting from $s_t$ after taking action $a_t$.

Notice that, finding $Q_*$ is equivalent to solving the MDP, since the optimal action for each state $s_t$ (hence the optimal policy $\pi_*$) can be obtained by 
\begin{align}
a^*_t = \argmax_{a_t} \; Q_*(s_t,a_t) .
\end{align}

{\em Relationships between $Q_*$ and $V_*$.}
From the definitions of $V_*(s_t)$ and $Q_*(s_t, a_t)$, it follows that
\begin{align}
V_*(s_t) = \max_{a_t} \; Q_*(s_t,a_t) ,  
\label{eq:VstarVsQstar}
\end{align}
i.e. the maximum cumulative reward from $s_t$ is the one corresponding to the maximum value of $Q_*$ after looking at all the potential alternative actions $a_t$. 
Conversely, we can write $Q_*(s_t, a_t)$ in terms of $V_*(s_t)$ as:
\small
\begin{align}
	Q_*(s_t,a_t) =  \sum_{s'_{t+1}} P_{a_t}(s_t,s'_{t+1}) (R_{a_t}(s_t,s'_{t+1}) + \gamma V_*(s'_{t+1})). 
	\label{eq:QstarVsVstar}
\end{align}
\normalsize
In other words, $Q_*(s_t,a_t)$ can be expressed as the immediate expected reward at time $t$, given by $\sum_{s'_{t+1}} P_{a_t}(s_t,s'_{t+1}) R_{a_t}(s_t,s'_{t+1})$, plus the expected cumulative reward from time $t + 1$ onwards, given by $\gamma \sum_{s'_{t+1}} P_{a_t}(s_t,s'_{t+1}) V_*(s'_{t+1})$.

Merging together equations~\eqref{eq:VstarVsQstar} and \eqref{eq:QstarVsVstar} we then obtain the Bellman Optimality Equation:
\begin{align} \label{eq:BellmanOptEq}
V_*(s) =  \max_a \Big\{ 
\sum_{s'} P_{a}(s,s') (R_a(s,s') + \gamma V_*(s')) \Big\} 
\end{align}

{\em Our strategy to solve the MDP.}
We have a complete description of our MDP (in particular, we have the transition probabilities $P_{a_t}(s_t,s'_{t+1})$ and the rewards $R_{a_t}(s_t,s'_{t+1})$), hence, in theory, we could enumerate all the possible states, use Dynamic Programming and the Value Iteration algorithm \cite{bellman_dynamic_1957} to find $V_*$ and calculate $Q_*$ via equation~\eqref{eq:QstarVsVstar}, thus solve the MDP. 
However, this is not a scalable approach due to the complexity of the MDP states and the very large number of successor states $s'$ for all but trivial networks. Instead, we use a Fitted Value Iteration algorithm \cite{Gordon_approximate_1999} that involves: 
$(a)$ devising a parametric representation $\bar{V}_*(s, \beta)$ for the optimal value function $V_*(s)$ in \eqref{eq:BellmanOptEq}, where $\beta$ is a placeholder for a set of parameters to fit (see section~\ref{ssec:ValueFunctApprox} and \ref{ssec:ValueFunctionParametrization}); 
$(b)$ using the approximate Bellman Optimality Equation (i.e. substituting $V_*(s')$ with $\bar{V}_*(s',\beta)$ in \eqref{eq:BellmanOptEq} 
to fit $\beta$ so that eventually $V_*(s) \approx \bar{V}_*(s, \beta^{fit})$ via a learning process (see section~\ref{ssec:learningProcess}); 
and finally $(c)$ calculating $Q_*(s,a)$ from $\bar{V}_*(s,\beta^{fit})$ hence solve the MDP.       


\subsubsection{Value function approximation}
\label{ssec:ValueFunctApprox}
In order to solve our MDP using the Fitted Value Iteration algorithm, we need a parametric representation of the optimal value function $V_*(s_t)$.
In our case, $V_*(s_t)$ is minus the minimum expected cumulative losses from $s_t$ (i.e. the maximum expected cumulative reward from state $s_t$) incurred between time $t$ to the end of the episode time $M$ (see also \eqref{eq:MDPRewardDefinition}-\eqref{eq:MDPVstarDefinition}). 
The greater the number of nodes in the financial network and the number of residual steps $m := M-t$, the greater is the potential for additional losses. 
It is natural to try to express $V_*(s_t)$ as a sum of the loss contributions due to each individual node at each of the remaining $m$ time steps. Hence, we introduce the matrix $\bar{Z}:=  (\bar{Z}_{ik}(s_t))$ with $i\in\I\setminus\I_{def}(t)$ and $k \in \{1,...,m\}$, 
where each element $\bar{Z}_{ik}(s_t)$ represents the approximate expected loss due to the default of node $i$ at time $t + k - 1$, taking into account potential government investments.    
Our final ansatz for the parametric representation $\bar{V}_*(s_t,\beta)$ of $V_*(s_t)$ is that it is given by a linear combination of the elements $\bar{Z}_{ik}$ in which the coefficients $\beta$ are arranged in a matrix that can change with time, i.e. $\beta \equiv \beta_t := (\beta_{ik}(t))$. Namely, 
\begin{align} \label{eq:ValueFunctionAnsatz}
       \bar{V}_*(s_t,\beta_t) &:=  - \sum_{i,k} \beta_{ik}(t) \bar{Z}_{ik}(s_t)  \\ \nonumber
       &\;\text{for } i\in\I\setminus\I_{def}(t), \;\; k \in\{1,..., m\}. 
\end{align}


We then let the system learn the parameters $\beta_{ik}(t)$ that maximise the expected cumulative reward (minimise the losses). 
In the following section, we describe how we fit the parameters $\beta_{ik}(t)$ to achieve the aforementioned task, while in section~\ref{ssec:ValueFunctionParametrization} we detail our choice of $\bar{Z}(s_t)$ in terms of the characteristics of the network.

\subsubsection{Learning process}
\label{ssec:learningProcess}
In order to learn the parameters $\beta_{ik}(t)$ we use the Bellman Optimality Equation \eqref{eq:BellmanOptEq}
and we define the ``Bellman value" as its right hand side after substituting $V_*(s')$ with the approximation $\bar{V}_*(s',\beta)$ from the previous section. Namely, we define 
\small
\begin{align} \label{eq:ValueBellDefinition}
&V_B(s_t, \beta_{t+1}) \\
&:= \max_{a_t} \Big\{ 
\sum_{s_{t+1}'} P_{a_t}(s_t,s_{t+1}') \big(R_{a_t}(s_t,s_{t+1}') + \gamma \bar{V}_*(s'_{t+1}, \beta_{t+1})\big) \Big\}. \nonumber
\end{align}
\normalsize
%
%
%

We can initialise $\beta$ with $\beta_{ik} (t) =1$ for all $i,k,t$, as a natural starting point due to our initial approximation of expected direct losses $\bar{Z}_{ik}(s_t)$ in \eqref{eq:ValueFunctionAnsatz} (see also Section~\ref{ssec:ValueFunctionParametrization} for more details).
We can then compare $\bar{V}_*(s_t, \beta_t)$ from \eqref{eq:ValueFunctionAnsatz} with $V_B(s_t, \beta_{t+1})$ from \eqref{eq:ValueBellDefinition} 
at state $s_t$ (starting from the initial state $s_0$ at time $0$ and moving forward to time $t$), and adjust $\beta$ so that the two values 
come closer. 
Afterwards, we move to another state 
$s'_{t+1}$ and repeat the same procedure until the difference between $\bar{V}_*$ and $V_B$ is ``small enough", 
within the subset of the state space $S$ that is reachable from $s_0$. 
Notice however, that the above approach does not converge in general, unless we use specific learning strategies.
The issue is that $V_B$ itself depends on $\beta$, which is what we want to fit, potentially triggering a divergent loop.
To resolve this issue, 
we notice that $\bar{V}_*(s_t,\beta_t)$ depends on $\beta$ at time $t$, while the corresponding $V_B(s_t,\beta_{t+1})$ is a function of $\beta$ at time $t+1$. 
Using this fact, if we fit $\beta$ backwards in time, then $\bar{V}_*(s_t, \beta_t)$ is compared with a value $V_B(s_t,\beta_{t+1})$ that is fixed (because $\beta_{t+1}$ would have been already fitted), thus solving the convergence problem. 

The primary issue that we now need to address is to 
find a way to calculate $V_B$ from \eqref{eq:ValueBellDefinition}, 
despite the fact that the set of states $s'$ that can be reached from state $s$ is huge, even for relatively small networks. 
We first notice that $\sum_{s'} P_{a}(s,s') R_a(s,s')$ is the `one-step' expected reward that can be rewritten in terms of the nodes of the network:
\begin{align}\label{Probs'i}
	\sum_{s'} P_{a}(s,s') R_a(s,s') = - \sum_{i\in\I\setminus\I_{def}} PD_i^a \; L_i^a  ,
\end{align}
with 
\begin{align}
	 PD_i^a  &:= PD(W_{i} + \Delta J^a_{i}, E_{i} + \Delta J^a_{i}, \mu_i, \sigma_i) ,\\ 
	 L_i^a   &:= \alpha_i \; (W_{i} +   \Delta J^a_{i})+ (J_i + \Delta J^a_{i}) \; LGD_i .
	 \label{eq:PDandLModifiedByAction}
\end{align}  
Secondly, the term $\sum_{s'} P_{a}(s,s') \; \bar{V}_*(s', \beta)$ can be estimated via Monte Carlo simulations, which involve $(a)$ sampling $s'$ using the distribution $P_a(s)$ defined by the probability mass function $P_a(s,s')$ and $(b)$ calculating the expected value $E^{P_{a}(s)}[\bar{V}_*(s', \beta)]$ by averaging the values $\bar{V}_*(s', \beta)$. 
Essentially, 
\begin{align}
	\sum_{s'} &P_{a}(s,s') \bar{V}_*(s', \beta) 
	\sim E^{P_{a}(s)}[\bar{V}_*(s', \beta)] .
	\label{eq:SecondTermVBMonteCarlo}
\end{align} 
However, it is not feasible to calculate $P_a(s,s')$ for all the states $s'$ that can be reached from $s$ after taking action $a$, due to the huge number of these states $s'$.
Once again, we use our knowledge of the underlying network dynamics to describe the right hand side of \eqref{eq:SecondTermVBMonteCarlo} in terms of nodes defaulting instead of MDP transition probabilities. We observe that the transition probability $P_a(s,s')$ was defined through the Gaussian latent variable model (see \eqref{eq:transProb}) and that there is a one-to-one correspondence between additional nodes defaulting from state $s$ and the state $s'$ reached given action $a$. 
In particular, we denote by $G^a_s$ the probability distribution
of states $s'$ which are derived by using our Gaussian latent variable model in order to first simulate which nodes default via the default mechanism in \eqref{eq:defaultCondition} and then to obtain the corresponding state $s'$. Since $G^a_s$ is equivalent to $P_a(s)$ due to the aforementioned one-to-one correspondence, we can therefore rewrite \eqref{eq:SecondTermVBMonteCarlo} as
\begin{align}
	\sum_{s'} &P_{a}(s,s') \bar{V}_*(s', \beta) 
	\sim E^{G^{a}_s}[\bar{V}_*(s', \beta)] .
	\label{eq:SecondTermVBMonteCarloGaussian}
\end{align} 
Putting all these together (using essentially \eqref{Probs'i} and \eqref{eq:SecondTermVBMonteCarloGaussian}) we can eventually rewrite $V_B(s_t, \beta_{t+1})$ from \eqref{eq:ValueBellDefinition} for all $t\in[0,M-1]$ in the form of
\small
\begin{align}\label{eq:VBellComplete}
V_B(s_t, \beta_{t+1}) = \max_{a_t} \bigg\{&-\sum_{i\in\I\setminus\I_{def}(t)} PD_i^{a_t} \ L_i^{a_t} 
\\\nonumber 
&+ \gamma \, E^{G^{a_t}_{s_t}} \big[\bar{V}_*(s'_{t+1}, \beta_{t+1}) \big] \bigg\}  
\end{align}
\normalsize
Now, given that our episode ends at time step $M$, we observe that $V_*(s_t) = 0$ for all $t \geq M$. 
Hence, at time $M-1$, we have that  $V_B(s_{M-1}, \beta_{M}) \equiv V_B(s_{M-1})$, since it will not depend on $\beta$, and we can thus write
\small
\begin{align}
V_B(s_{M-1}) &=  \max_{a_{M-1}} \bigg\{- \sum_{i\in\I\setminus\I_{def}(M-1)} PD_i^{a_{M-1}} L_i^{a_{M-1}} \bigg\} \nonumber\\
&= V_*(s_{M-1}),
   	\label{eq:VBellMinus1}
\end{align} 
\normalsize
where the latter equality follows from \eqref{eq:BellmanOptEq} and \eqref{Probs'i}.
Now that we can calculate the exact optimal value function $V_*$ for each state at time $M-1$, we notice from \eqref{eq:VBellComplete} that $V_B(s_{M-2}, \beta_{M-1}) \equiv V_B(s_{M-2})$ is also independent of $\beta$, namely
%
\small
\begin{align}
V_B(s_{M-2}) = \max_{a_{M-2}} \bigg\{ &-\sum_{i\in\I\setminus\I_{def}(M-2)} PD_i^{a_{M-2}} \; L_i^{a_{M-2}} \nonumber\\
                                      &+ \gamma \, E^{G^{a_{M-2}}_{s_{M-2}}}  \big[V_*(s'_{M-1})\big] \bigg\} .
\end{align}
\normalsize
We then fit $\beta$ backwards in time for the decreasing sequence of time steps $(M-2, ..., 0)$, creating a representative portfolio of MDP states for each time step (see section~\ref{ssec:RapresentativePortfolio}) and performing a ridge regression (with a 5-fold cross-validation) comparing $\bar{V}_*(s_t, \beta_t)$ with $V_B(s_t, \beta_{t+1})$. 
Firstly, for time step $M-2$, we compare $\bar{V}_*(s_{M-2},\beta_{M-2})$ with  $V_B(s_{M-2})$, for all the states in the representative portfolio, and we fit $\beta_{M-2}$. 
Then, for time step $M-3$, we calculate
\small
\begin{align}
V_B(s_{M-3}, \beta_{M-2}) &=  \max_{a_{M-3}} \bigg\{-\sum_{i\in\I\setminus\I_{def}(M-3)} PD_i^{a_{M-3}} \; L_i^{a_{M-3}}  \nonumber\\
                          &+ \gamma \, E^{G^{a_{M-3}}_{s_{M-3}}}[\bar{V}_*(s'_{M-2}, \beta_{M-2})]\bigg\} 
\end{align}
\normalsize
and compare it with $\bar{V}_*(s_{M-3}, \beta_{M-3})$, for all the states of the representative portfolio, to obtain once again  $\beta_{M-3}$ via a ridge regression.
We continue the procedure backward in time until we successfully obtain $\beta^{fit}$, i.e. the fitted $\beta_t$ for each time $t$. 

\subsubsection{Solution of MDP}

Finally, we can solve the MDP by combining all the above results to calculate $Q_*(s_t, \alpha_t)$. In particular, by  substituting $V_*(s'_{t+1})$ in \eqref{eq:QstarVsVstar} with its approximation $\bar{V}_*(s'_{t+1}, \beta^{fit}_{t+1})$ obtained in Section \ref{ssec:learningProcess}, and applying the same analysis performed to obtain \eqref{eq:VBellComplete}, we conclude that
\begin{align}\label{eq:QstarWithBetaFit}
Q_*(s_t, \alpha_t) \approx  \bigg\{&-\sum_{i\in\I\setminus\I_{def}(t)} PD_i^{a_t} \ L_i^{a_t} 
\nonumber\\ 
&+ \gamma \, E^{G^{a_t}_{s_t}} \big[\bar{V}_*(s'_{t+1}, \beta^{fit}_{t+1}) \big] \bigg\},  
\end{align}
which provides the solution to the MDP.


\section {Results}
\label{sec:Results}
\begin{figure}[ht!]
	\centering
	\includegraphics[width=85mm]{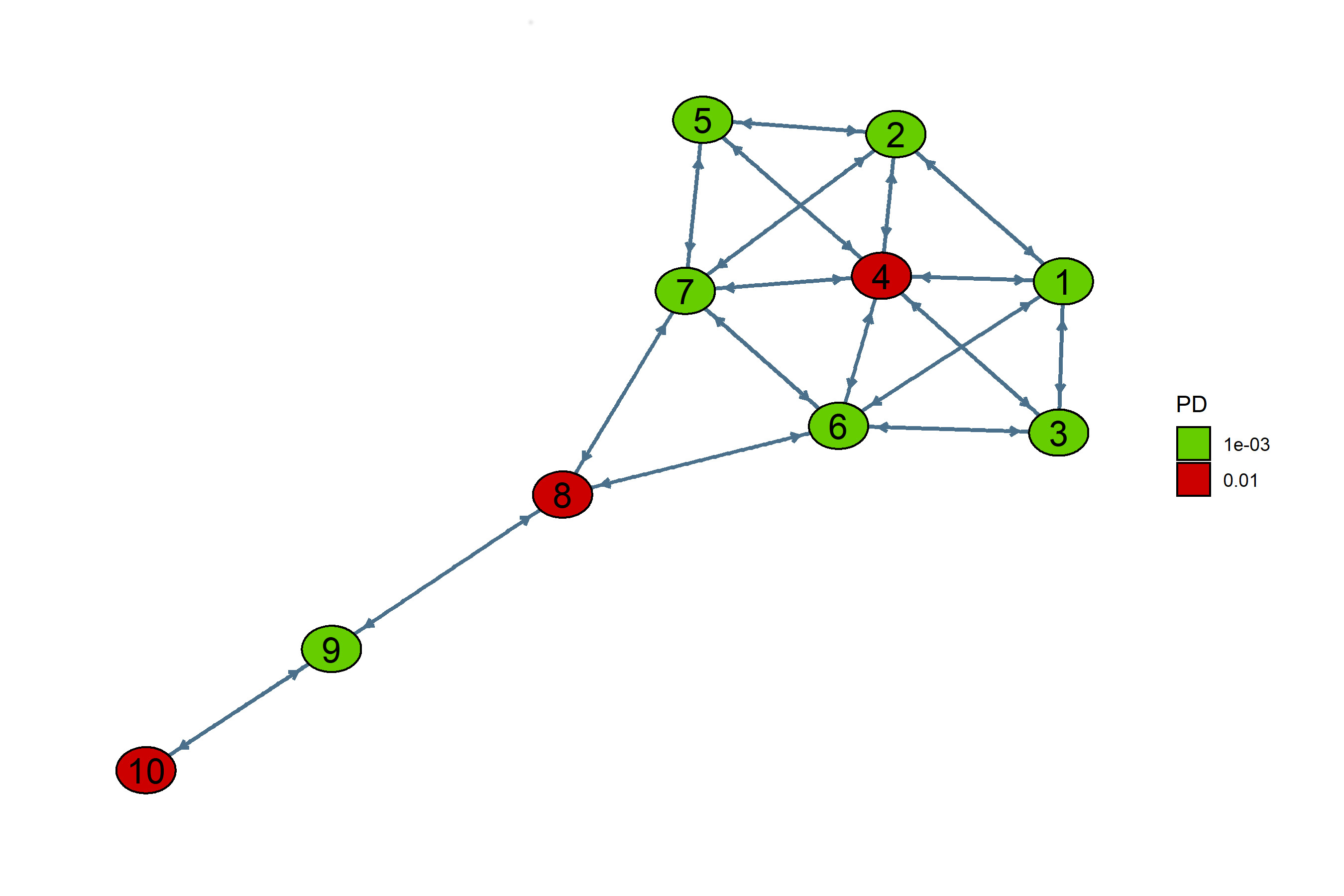} 
	\caption{Krackhardt kite (KK) graph. In our example we consider:
	 set of nodes $\I=\{1,...,10\}$, total asset $W_i(0) = 100$, capital $E_i(0) = 3$, $\mu_i = 0$, $LGD_i = 1$, for all $i\in\I$, $PD_i(0) = 0.01$ for $i\in\{4,8,10\}$ and $PD_i(0) = 0.001$ for $i\in\I\setminus\{4,8,10\}$}
	\label{fig:KKpgraph}
\end{figure}

The main result of this paper is the creation of the framework itself. A professional calibration of our model would require the effort of a central bank or a government office. To show how our model works, we explore two instances of our framework: 
In Section \ref{KKn} we use a network with homogeneous nodes organised as the Krackhardt kite \cite{Krackhardt_1990} (KK) graph (Fig.~\ref{fig:KKpgraph}), while 
in Section \ref{GSIIn} we use the network of the European Global Systemically Important Institutions (GSIIs) obtained from the data in the European Banking Authority website \footnote{https://eba.europa.eu/} (EBA network).  

\subsection{Krackhardt kite network} \label{KKn}

We assume the number of steps to be $M = 7$, the discount factor $\gamma = 0.98$ and for each node $i$, $J_i = 0$ (unless otherwise specified), $\mu_i = 0$ (assuming conservatively that the expected value of the assets' return is zero), $\alpha_i = \alpha$ (the same for each node) and that the government can invest only in ``risky" banks $i$ with ``relatively high" $PD_i$ (in our examples, ``risky" banks will have $PD_i > 0.009$). 
We have used an homogeneous correlation matrix that takes into account the average correlation between banks and following \cite{huang_framework_2009} we have set $\Sigma_{ij} = 0.5$ for $i \ne j \in \I\setminus\I_{def}$.  
The value of $\sigma_i$, for each node $i$, is calculated at time $t=0$ from $PD_i(0)$, $E_i(0)$ and $W_i(0)$, by inverting \eqref{eq:PDMLongVersion}. 
Finally, we set $PDM^{Floor}_i = 0.00021$ which is the upper end of the AAA default probability bracket, within the internal credit rating methodology used by Credit Suisse~\cite{CreditSuisse_2019}. 
The available actions are expressed with the notation: $<$node$>@<$capital investment as a tenth of a percent of the total asset W$>$. For example, 8@05 means an investment of $50\, bp \; W_8$ or $0.5\% \, W_8$ 
in node 8. An action that considers all the nodes is indicated with $<$node$>=0$. Hence 0@15 stands for an investment of $1.5\% \, W_i$ in each ``risky" 
node $i\in\I\setminus\I_{def}$.  

The common theme is that adding external resources makes the network more resilient but they can be lost in a subsequent default, which creates a trade-off for the decision maker. 
For relatively low values of $\alpha$, it is generally not convenient to invest, while for relatively high values of $\alpha$, the best action is to invest an amount of capital that makes the network sufficiently resilient. 
In the EBA network (see Section \ref{GSIIn}), we have shown that there exists a ``critical" $\alpha_c$ that 
splits the space of $\alpha$-values into two ``regimes" of low/high values, where $\alpha_c \approx 0.0025$ in the original network and $\alpha_c \approx 0.00096$ in the distressed network where the capital of the banks was halved.

\begin{figure}[ht!]
	\centering
	\includegraphics[width=80mm]{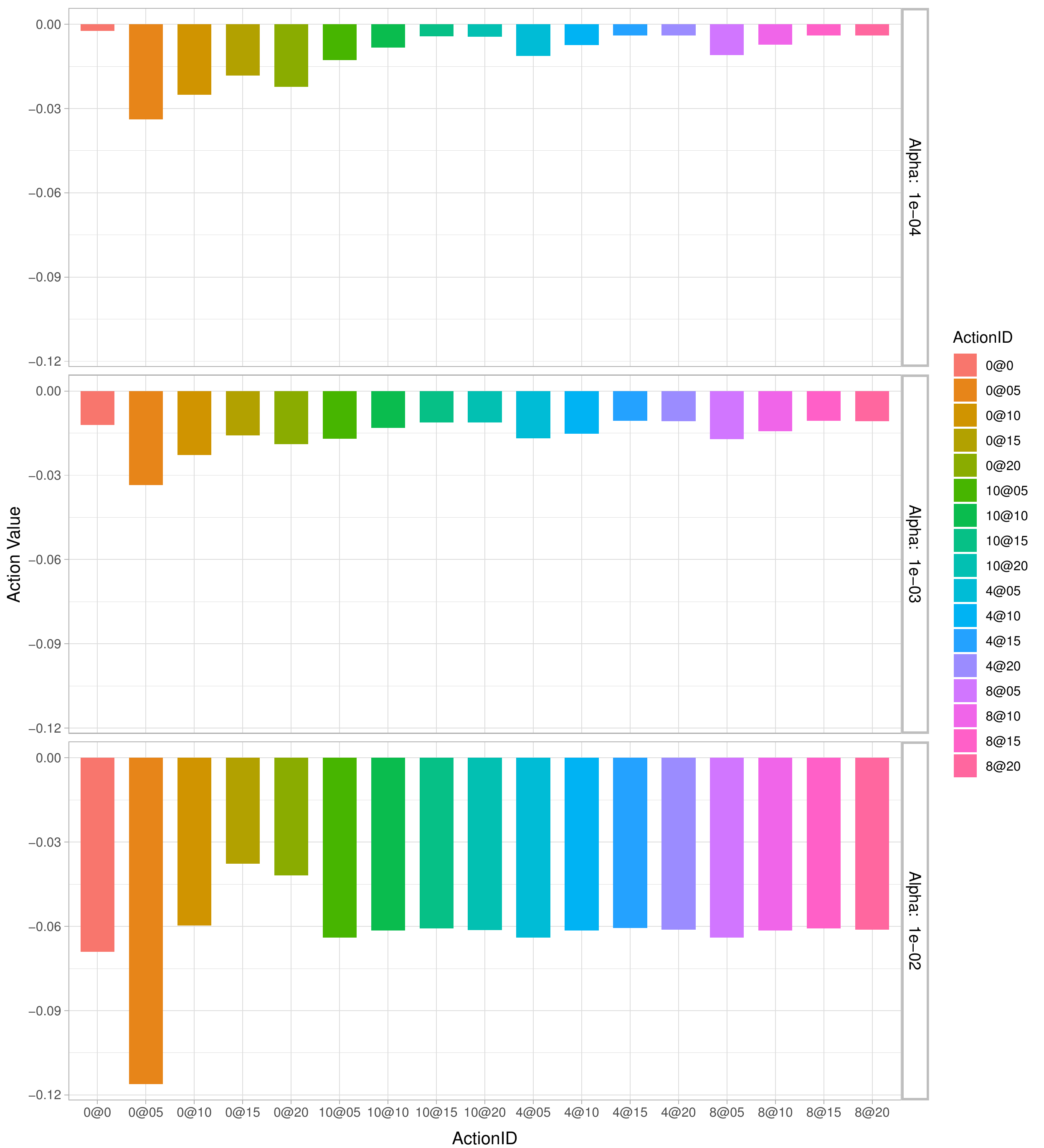} 
	\caption{The picture is relative to the 
	KK network and shows the optimal action value at time $t = 0$ for different actions and values of $\alpha$. In the legend, 0@0 means no investment, 0@05 means investing 0.5 in all the nodes with $PD > 0.009$ (i.e nodes 4, 8, 10), 10@05 means investing 0.5 in node 10, 4@10 means investing 1 in node 4, etc.
	It is never convenient investing the maximum amount of capital (0@20): 2 for each node with $PD > 0.009$ (i.e nodes 4, 8, 10). For small values of alpha, the best action is not to invest (0@0), as alpha increases, so does the convenience of investing more capital. For $alpha = 1e-02$ the best action is to invest 1.5 in nodes 4, 8 and 10 (0@15). } 
	\label{fig:KKALLBARw10}
\end{figure}

We have chosen this particular network to assess if our algorithm can distinguish between central nodes and peripheral ones. All the nodes have $W(0) = 100$, $E(0) = 3$, $\mu = 0$, $LGD = 1$ (we assume, conservatively, that all the investment would be lost in case of a default),  $PD(0) = 0.001$ for all but nodes 4, 8 and 10 with $PD(0) = 0.01$.
The edges between nodes are oriented and homogeneous, assuming the value $w_{ij} = 1 \; \forall  i\not=j$.
We have restricted the potential investment amounts, for each node,  to be: $0$, $0.5\% \, W$, $1\% \, W$, $1.5\% \, W$ or $2\% W$. Furthermore, the government can choose to invest in a single node or all the nodes for each time step, provided that the nodes are considered distressed. In our example, a node $i$ is defined as risky or distressed if $PD_i > 0.009$.

\begin{figure}[H]
	\centering
	\includegraphics[width=75mm]{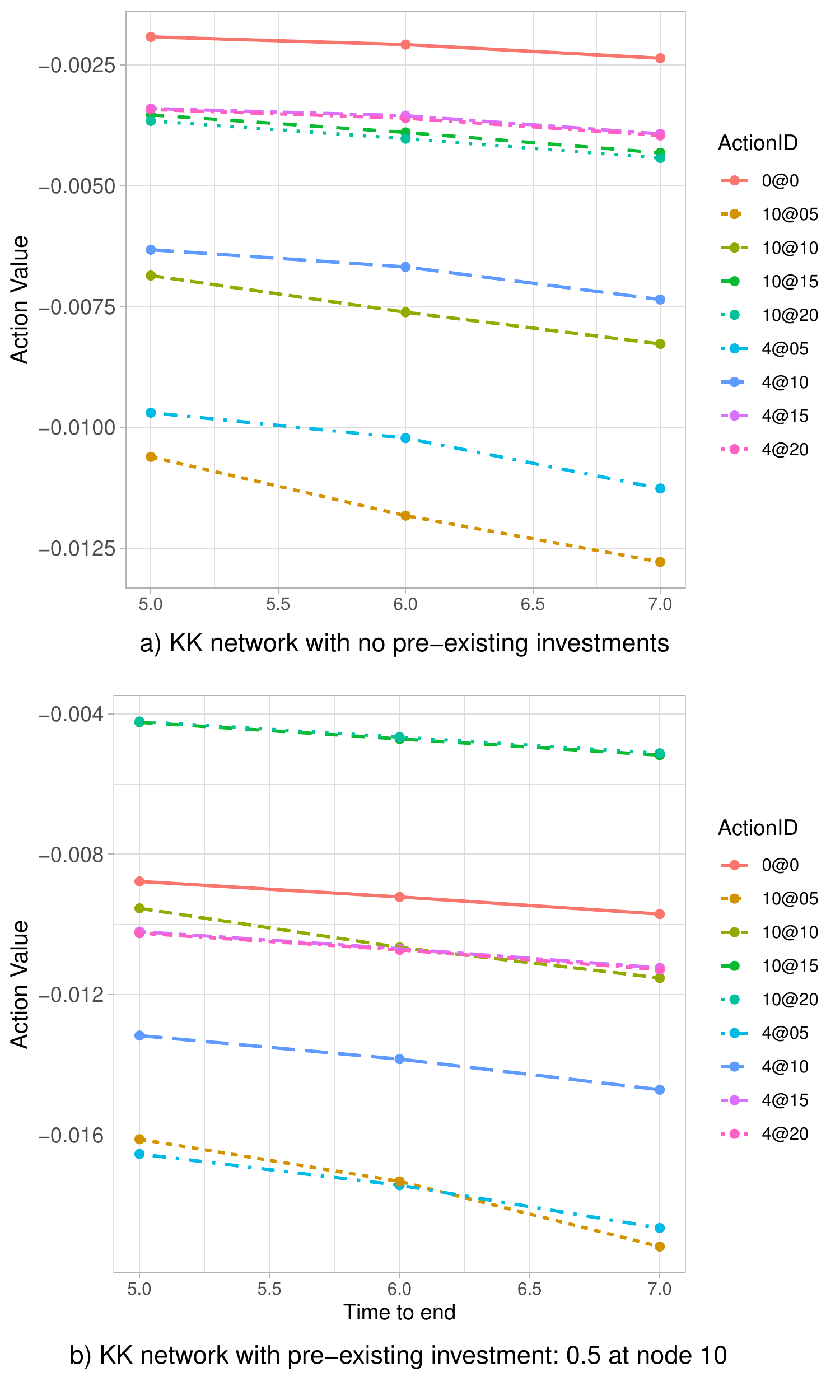} 
	\caption{The pictures are relative to the Krackhardt kite graph (KK) and shows the optimal action value vs time to end of the episode, with $alpha = 0.0001$, for different actions, and focusing on nodes 4 (central node) and 10 (peripheral node).
	(a) The algorithm `feels' the network structure and suggests to invest in node 4 rather than node 10. 
	(b) In case the government had previously invested in node 10, the government needs to protect its investment, risking an additional investment in node 10. In the legend, 0@0 means no investment, 10@05 means investing 0.5 in node 10, 4@10 means investing 1 in node 4, etc. } 
	\label{fig:KKQvsTimeToEnd}
\end{figure}

We have analysed the system for different values of alpha (0.0001, 0.001, 0.01) and reported the optimal action values at time $t = 0$ in (Fig.~\ref{fig:KKALLBARw10}). 
For $\alpha = 0.0001$ the best action is 0@0 (i.e. no investment in any node) followed by investing the minimum amount of capital in individual nodes. As alpha increases, not to invest becomes less and less convenient compared to the other options. For $alpha = 0.01$ the best action is to invest 1.5 in all the risky nodes (0@15). 
It is interesting to note that action 0@2 (i.e. investing the maximum amount, 2, in all the risky nodes) is never the best choice, while 0@05 is always the worst, as it provides too few capital to each node to make them resilient. In Fig.~\ref{fig:KKQvsTimeToEnd}(a) we can see that the optimal action values corresponding to investments in different nodes tend to converge as the time to the end of the episode decreases because the contagion has less time to propagate and the node position becomes less and less relevant.

In Fig.~\ref{fig:KKQvsTimeToEnd}(a), we focus our analysis on nodes 4 and 10 for $\alpha = 0.0001$ and we note that investing in node 4 is always better than in node 10 for the same amount of capital.
In Fig.~\ref{fig:KKQvsTimeToEnd}(b), we show how the results would change in the case when the government had already invested 0.5 in node 10 (i.e. $J_{10}(0) = 0.5$). In this case, a substantial  investment in node 10 (10@15 or 10@20) largely outperform investments in node 4. The government needs to keep investing a sufficient amount of capital in node 10 to protect its previous investment. For example, an investment of 0.5 in node 10 is not sufficient to strengthen it and the corresponding action value is the worst among the one considered at time $t = 0$  (time to end = 7).

\subsection{European GSII network} \label{GSIIn}

We use the data from the European Banking Authority website about the Global Systemically Important Institutions, relative to the year 2014 (EBA network) \cite{EBA_2014}. The data does not contain the complete bilateral network (as this is considered business sensitive information) but aggregates of credit exposures vs other financial institutions. For our analysis, we have used the algorithm described in our previous paper \cite{petroneLatora_dynamic_2018} (see also \cite{anand_filling_2015}) to reconstruct the network. 

\small
\begin{table}[b]
	\centering
			\begin{tabular}{lrrrll}
				\noalign{\smallskip} \hline \hline \noalign{\smallskip} 
                    SYMBOL&W&E&PD&BANK&\\
                    \hline
                    BFA&235&12&0.0116&BFA&\\
                    MPS&201&7&0.0093&Monte dei Paschi di Siena&\\
                    UNI&1034&45&0.0017&Unicredit&\\
                    INT&696&38&0.0017&Intesa Sanpaolo&\\
                    CAI&377&19&0.0017&La Caixa&\\
                    BNP&2253&70&0.001&BNP Paribas&\\
                    BAR&1940&59&0.001&Barclays&\\
                    CAG&1723&71&0.001&Credit Agricole&\\
                    DEB&1659&63&0.001&Deutsche Bank&\\
                    SAN&1456&64&0.001&Santander&\\
                    RBS&1411&51&0.001&RBS&\\
                    SOC&1409&47&0.001&Societe Generale&\\
                    BPC&1337&50&0.001&BPCE&\\
                    ING&1164&41&0.001&ING&\\
                    LOY&1107&46&0.001&Lloyds&\\
                    BBV&723&42&0.001&BBVA&\\
                    CMU&695&37&0.001&Credit Mutuel&\\
                    COM&656&25&0.001&Commerzbank&\\
                    DAN&494&19&0.001&Danske Bank&\\
                    ABN&421&16&0.001&ABN Amro&\\
                    DZB&356&13&0.001&DZ Bank&\\
                    DNB&332&15&0.001&DNB&\\
                    SEB&310&13&0.001&SEB&\\
                    LBW&290&13&0.001&LBBW&\\
                    BLB&275&10&0.001&Bayern LB&\\
                    SWE&249&10&0.001&Swedbank&\\
                    KBC&232&14&0.001&KBC&\\
                    POS&223&7&0.001&Banque Postale&\\
                    ERS&219&11&0.001&Erste Group&\\
                    NLB&216&7&0.001&NordLB&\\
                    HLB&199&8&0.001&Helaba&\\
                    HSB&2680&117&0.0004&HSBC&\\
                    RAB&728&34&0.0004&Rabobank&\\
                    NOR&655&25&0.0004&Nordea&\\
                    HAN&334&11&0.0004&Handelsbanken&\\
				\noalign{\smallskip} \hline \noalign{\smallskip}
			\end{tabular}
			
	\caption{European Union Global Systemically Important Institution (GSII). Tier 1 capital (E) and  total asset (W) are expressed in billion of EUR. The data are from the European Banking Authority (EBA) website and are relative to the end of 2014. The probabilities of default have been derived using data from the credit rating agency Fitch (at www.fitchratings.com). The values of W and E are expressed in billion of EUR.} 
	\label{t:EBA_table1}
\end{table}
\normalsize

\begin{figure}[b]
	\centering
	\includegraphics[width=85mm]{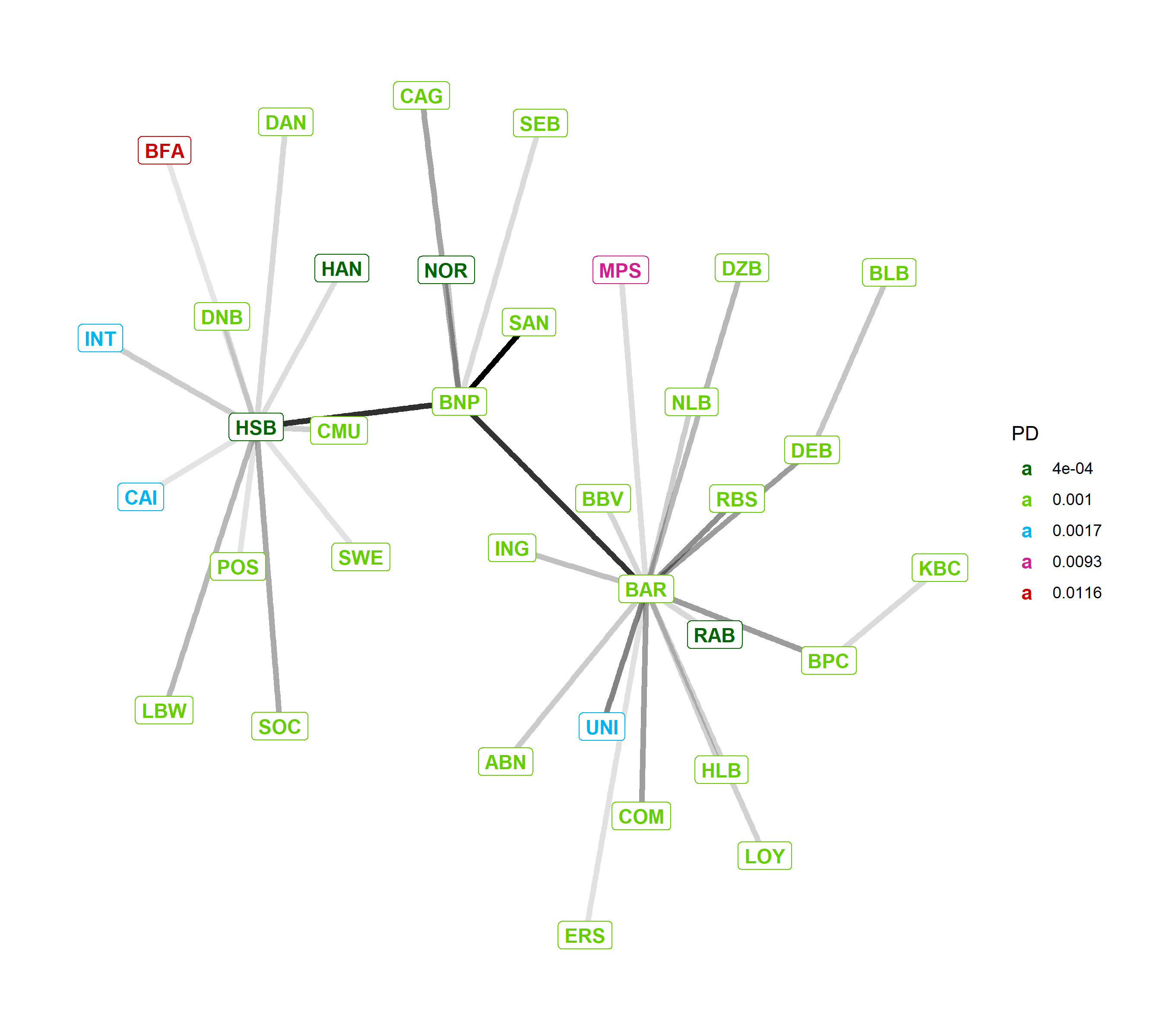} 
	\caption{Maximum spanning tree of the EBA graph. Each node represents a financial institution (see Table~\ref{t:EBA_table1}). The graph as been reconstructed from aggregated data available at the European Banking Authority (EBA) website and it can be different from the actual network of bilateral exposures. The darker edges identify stronger exposures. The nodes with higher PD are Monte dei Paschi di Siena (MPS) and BFA.}
	\label{fig:EBAgraph}
\end{figure}

\begin{figure}[ht!]
	\centering
	\includegraphics[width=80mm]{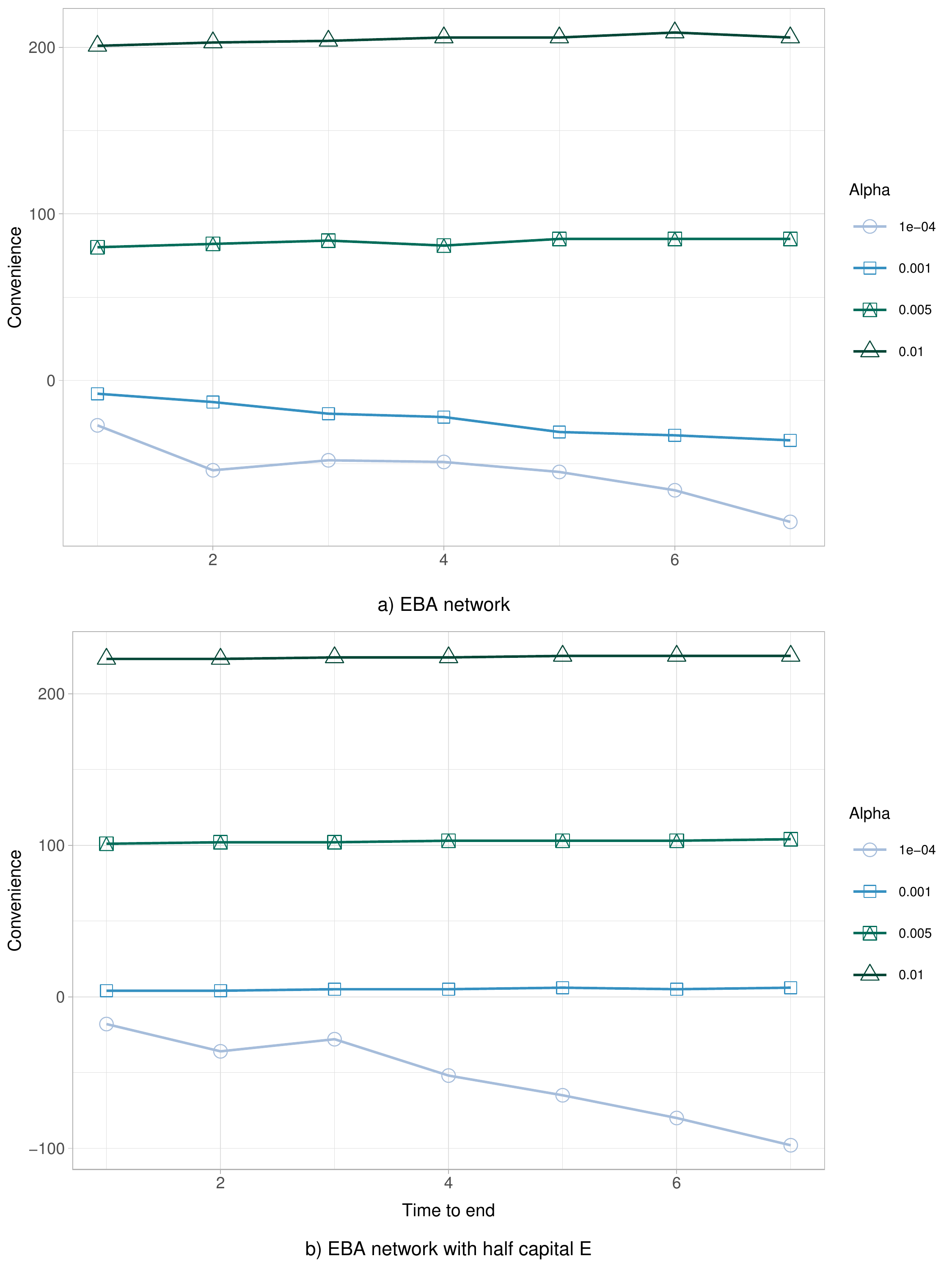} 
	\caption{(a) The Convenience, expressed in million of EUR in the charts, and defined as the difference between the optimal action value corresponding to the best government intervention, and the optimal action value associated to inaction (see \eqref{eq:Convenience}),  is almost constant vs time to the end of the episode for positive values, and a decreasing function for negative values. 
	(b) If we stress the EBA network, halving the capital of the nodes, we obtain a chart similar to a) but the minimum value of $alpha$ for which the Convenience is positive is lower.} 
	\label{fig:EBAStdAndHEConvenienceVsTimeToEnd}
\end{figure}

We set $LGD = 0.6$ as this is the standard rule of thumb in financial credit risk \cite{altman_default_2004}. The values of the asset volatility $\sigma$ have been obtained inverting \eqref{eq:PDMLongVersion} at time zero and assuming it remains constant during the simulation.
We have restricted the potential investment amounts to be: $0$, $0.5\% \, W$, $1\% \, W$, $1.5\% \, W$,  $2\% W$, $2.5\% W$, $3\% W$. Furthermore, if the government decides to invest, it needs to provide additional capital to all the risky nodes, defined in our example, as the nodes with $PD > 0.009$. 
For this exercise, we pretend that the European Union (including UK) is also a fiscal union with a single government that is accountable to all the European taxpayers. In particular, we consider investments that individual states might have in banks as of 2014 as `private' investments, hence we start with with $J_i(0) = 0 \; \forall i$.
We define the `Convenience' to intervene as:
\small
\begin{align}
Convenience(s_t) := \max_{a_t \ne a^0_t} \{Q_*(s_t, a_t)\} -  Q_*(s_t, a^0_t)
\label{eq:Convenience}
\end{align}
\normalsize
with $a^0_t$ representing the action at time $t$ corresponding to no investments.
In Fig.~\ref{fig:EBAStdAndHEConvenienceVsTimeToEnd}(a) we have reported the Convenience vs the time (number of steps) to the end of the episode.
We have found that the Convenience is positive and almost constant for large values of $\alpha$ ($\alpha = 0.01$, $\alpha = 0.005$), and is negative and decreasing for smaller values of $\alpha$ ($\alpha = 0.001$, $\alpha = 0.0001$). 
In Fig.~\ref{fig:EBAStdAndHEConvenienceVsTimeToEnd}(b) we have the same chart but for a severely distressed version of the network, where the capital of the banks has been halved. The distress has the effect of lowering the value of alpha at which the Convenience is positive, for example the Convenience is now positive for $\alpha = 0.001$.

\begin{figure}[H]
	\centering
	\includegraphics[width=80mm]{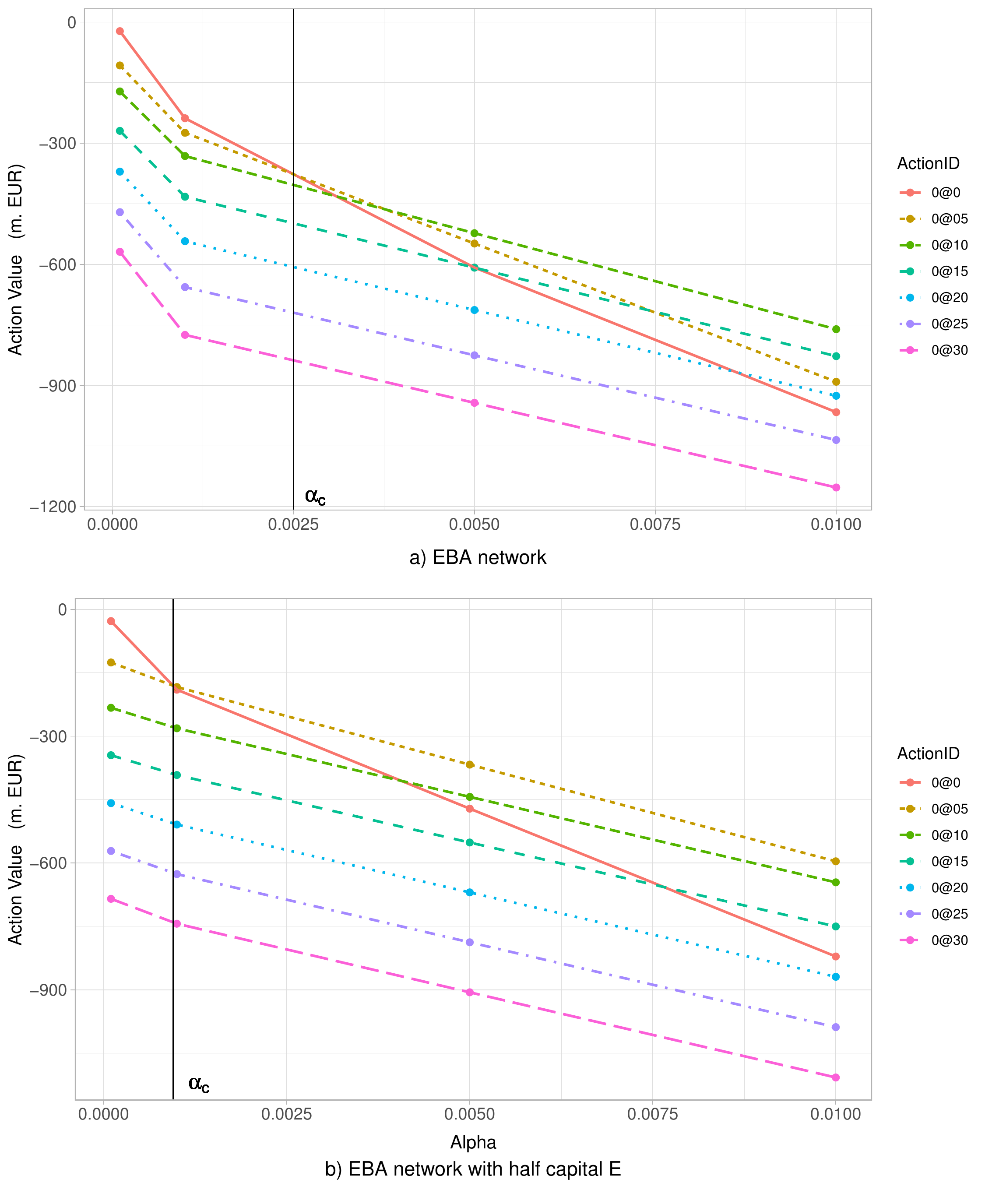} 
	\caption{a) The chart is relative to the `EBA network' (European Global Systemically Important Institutions). The optimal action value at time $t = 0$ is reported as a function of $\alpha$ for different actions. 0@0 means no investment. As $\alpha$ increases, 0@0 becomes less and less convenient and for $\alpha = \alpha_c = 0.0025$ the best action becomes investing $0.5\% W$ in each of the risky nodes. For even higher values of $alpha$, the best action becomes 0@10 (i.e investing $1\% W$ for each of the risky nodes).
	b) The chart is relative to a distressed version of the `EBA network' where the capital of the banks has been halved. The value of $\alpha_c$ at which a government intervention is convenient is lower than in a):
	 $\alpha_c \approx 0.00096$.} 
	\label{fig:EBAStdAndHEQvsAlphaByActionID}
\end{figure}

To explore the transition between positive and negative Convenience, we have reported the optimal action values at time $t = 0$ as a function of $\alpha$ in (Fig.~\ref{fig:EBAStdAndHEQvsAlphaByActionID}(a)). 

For $\alpha > \alpha_c \approx 0.0025$, the inaction is no longer the most convenient choice, ``0@05" (investing $0.5\% W$ in the risky nodes) becomes the best action. It is also interesting to notice that the optimal action becomes ``0@10" for higher values of $\alpha$.       
In (Fig.~\ref{fig:EBAStdAndHEQvsAlphaByActionID}b)) we have the optimal action values at time $t = 0$ when the capital of the banks has been halved. We notice that the value $\alpha_c$ at which a government intervention becomes favourable is lower at $\alpha_c \approx 0.00096$. 

\section{CONCLUDING REMARKS}
\label{sec:ConcludingRemarks}
We have shown how to cast a bank bailout decision by a government into an action in a Markov Decision Process (MDP) where the states of the MDP are defined in terms of the underlying network of financial exposures and the MDP dynamics is derived from the network dynamics.
In our example, that uses the data relative to the European Global Systemically Important Institutions, we have found that government interventions do not improve the expected loss of the financial network if the loss for the taxpayer as a fraction of the bank total assets $\alpha$ satisfies $\alpha < \alpha_c \approx 0.0025$. The value of $\alpha_c$ becomes lower as the distress of the network increases. 
It is evident from our analysis that the parameter $alpha$ plays a central role in systemic risk modelling and even if there are works \cite{Blix_2016} \cite{Cariboni_reducing_2016} studying the impact for the taxpayers linked to a bank default, additional analysis need to be performed for its reliable estimation.
Using a simplified Krackhardt kite network, we have found that the government becomes biased toward investing in a risky node if it had already invested in it in the past. The government needs to evaluate carefully a potential investment. The rescued bank could increase its risky investments knowing that it would be bailed-out in case it became distressed again, thus leading to moral hazard. 

\section {Material and Methods}
\subsection{Representative portfolio of MDP states}
\label{ssec:RapresentativePortfolio}
In section~\ref{ssec:ValueFunctApprox} we have expressed the approximated value function $\bar{V}_*(s_t, \beta)$  as a linear combination of terms $\bar{Z}_{ik}(s_t)$ with coefficients $\beta_{ik}(t)$ given by \eqref{eq:ValueFunctionAnsatz}.%

In order to fit these $\beta_{ik}(t)$, we first identify a representative portfolio of MDP states that can be reached, at time $t$, from the initial state $s_0$, and for which we can calculate the Bellman value $V_B$ using \eqref{eq:VBellComplete}--\eqref{eq:VBellMinus1}. Equating $\bar{V}_*(s_t, \beta_t)$ with the corresponding $V_B(s_t, \beta_{t+1})$, for each state $s_t$ in the portfolio, we derive a set of linear equations that we use to obtain the coefficients $\beta_{ik}(t)$ via a ridge regression (with a 5-fold cross-validation).
The states in the representative portfolio, at time t, are obtained from the initial state $s_0$, after changing the time to maturity from $M$ to $M - t$ (i.e. the states are `moved' forward in time) and forcing a set $U$ of nodes to default. The representative portfolio contains: 
$(a)$ the state corresponding to $U = \emptyset$, 
plus 
$(b)$ all the states corresponding to $U = \{i\}$ for $i \in \I$ (i.e. with one additional defaulted node with respect to $s_0$), plus 
$(c)$ a selection of states corresponding to $|U| > 1$ (i.e. with multiple additional defaulted nodes), which are chosen randomly with probabilities proportional to $exp \, (-|U|)$ (a greater importance is given to states with fewer number of additional defaults as they are more likely to be reached in an actual simulation).
In addition, we obtain elements in the representative portfolio by performing a government action on $s_0$ and then move the corresponding state at time $t$ (i.e. at time to maturity $M-t$). 
The number of states in the representative portfolio needs to be chosen taking into account the trade-off between stable results and computational resources.

\subsection{Value function parametrisation}
\label{ssec:ValueFunctionParametrization}
In this section, we detail our choice for $(\bar{Z}_{ik}(s_t))$ used in our ansatz for the value function approximation $\bar{V}_*(s_t, \beta)$ in \eqref{eq:ValueFunctionAnsatz}, with node $i\in\I\setminus\I_{def}(t)$ and step $k \in\; \{1,...,m = M-t\}$ until the end of the episode.
We introduce the auxiliary matrix $Z$, with elements $Z_{ik}(s_t;a_1,...,a_k)$ representing the approximated contribution of the expected direct loss, due to the default of node $i$, at time $t+k-1$, taking into account the government actions $a_j$ at time $t+j-1$, for all $j=1,...,k$. 
That is, we define  
\begin{align}
		\label{eq:Zik}
		    Z_{ik} &:=
		    \begin{cases}
		    	PD_{ik} \; L_{ik} , & \text{if}\ k=1 ; \\
		    	PD_{ik} \; L_{ik}  \; \gamma^{k -1}  \prod_{r=1}^{k-1} (1 - PD_{ir}), & \text{if}\ k>1.
		    \end{cases} \nonumber
\end{align}
Here, the value $PD_{ik}$ is the modified probability of default and the value $L_{ik}$ is the modified loss, associated to the node $i$ at time $t+k-1$, that take into account the expected cumulative impact $I_{ik}$ and the potential cumulative investment from the government $J_{ik}$ on node $i$ from time $t$ up to time $t+k-1$. 
Note that for $k>1$, a node $i$ can contribute to the expected loss only if it has not defaulted in the previous time steps (hence the presence of the survival probabilities $(1 - PD_{ir})$).
To be more precise, we firstly define
\small
\begin{align}
\nonumber
&PD_{ik} \\ 
&:= PD(W_{i} + J_{ik} - I_{ik}, E_{i} + J_{ik} - I_{ik}, \mu_i, \sigma_i, PDM^{floor}_i). 
\nonumber
\end{align}	
\normalsize
Then, the cumulative impact $I_{ik}$ depends on the modified probability of default of all the nodes $j \in {H}:=\I\setminus(\I_{def}(t) \cup \{i\})$ 
and is defined by
\begin{align}
	\nonumber
 		I_{ik} &:=
 		\begin{cases}
 			0 & \text{if}\ k=1; \\
 			\sum_{j \in H}  PD_{j1} w_{ij}   & \text{if}\ k = 2; \\
 			I_{i \, k - 1} + \sum_{j \in H}  PD_{j \, k - 1} w_{ij}  \prod_{r=1}^{k-2} (1 &\hspace{-10pt}- PD_{jr}) \\ 
 			&\text{if}\ k > 2. \\
 		\end{cases} 
\end{align}
Moreover, the cumulative government investment $J_{ik}$ in node $i$ is a function of the actions $(a_1, ..., a_k)$ that the government can take between $t$ and $t+k-1$ and is defined by 
\begin{align} \nonumber
J_{ik} ( a_1,...,a_k ) :=   \sum_{r=1}^{k} \Delta J^{a_r}_i(t+r-1)  
\end{align}
Finally, the modified loss incurred is defined by  
\begin{align} \nonumber
&L_{ik} :=  \alpha_i \; (W_{i} +   J_{ik} - I_{ik})+ (J_i + J_{ik}) \; LGD_i 
\end{align}
In light of the above equations, we observe that $Z_{ik} = Z_{ik}(s_t; a_1,...,a_k)$ depend on the actions $(a_1,...,a_k)$ via the terms  $J_{ik}(a_1,...,a_k)$ involved in both $PD_{ik}$ and $L_{ik}$.

We now call $a^0$ the action corresponding to no additional government investment and define the total expected direct loss for all $i\in\I\setminus\I_{def}(t)$ and $k \in \{1,...,m\}$ as
\begin{align}
\nonumber
TL(s_t;a_1,a_2,..,a_m) 
&:= \sum_{i,k} Z_{ik}(s_t;a_1,a_2,..,a_k ) . 
\end{align}
Then, the specific matrix $\bar{Z} = (\bar{Z}_{ik}(s_t))$ involved in our value function approximation is defined by 
\begin{align}
	\nonumber
	\bar{Z}_{ik}(s_t) := Z_{ik}(s_t;\bar{a}_1,...,\bar{a}_k), 
\end{align}
where each $\bar{a}_{j}$ is calculated sequentially for each $j \in \{1, ..., m\}$ as follows: 
\begin{align}
	\nonumber
	\bar{a}_{1} &:= \argmin_{a_1} TL(s_t;a_1,a_2^0,a_3^0,..,a^0_m) \\   \nonumber
	\bar{a}_{2} &:= \argmin_{a_2} TL(s_t;\bar{a}_1,a_2, a_3^0,..,a^0_m ) \\ \nonumber
	& \qquad \qquad \vdots \\
	\bar{a}_{m} &:= \argmin_{a_m} TL(s_t;\bar{a}_1, \bar{a}_2, ..., \bar{a}_{m-1}, a_m) . \nonumber
\end{align}



\bibliographystyle{amsplain}



\providecommand{\bysame}{\leavevmode\hbox to3em{\hrulefill}\thinspace}
\providecommand{\MR}{\relax\ifhmode\unskip\space\fi MR }
\providecommand{\MRhref}[2]{%
	\href{http://www.ams.org/mathscinet-getitem?mr=#1}{#2}
}
\providecommand{\href}[2]{#2}

\bibliographystyle{amsplain}


\bibliography{zot_Shocks29Sep16}

\begin{thebibliography}{10}
	
	\bibitem{Krackhardt_1990}
	Krackhardt,~D. Assessing the Political Landscape: Structure, Cognition, and Power in Organizations.
	\emph{Adm. Sci. Q.} \textbf{35(2)}, 342--369, (1990)
	
	\bibitem{BOR_EFO_2018}
	Office~for~Budget~Responsibility. Available at
	http://cdn.obr.uk/EFO-MaRch\_2018.pdf\#page=106 (2018).	
	
	\bibitem{NAO_Managing_2009}
	National~Audit~Office. Available at 
	https://www.nao.org.uk/wp-content/uploads/2009/12/091091.pdf (2009)
	
	\bibitem{EBA_2014}
	National~Audit~Office. Available at https://eba.europa.eu/risk-analysis-and-data/global-systemically-important-institutions,2014 data (2015)
	
	\bibitem{RBS_ShareOwnership_2020}
	Natwest~Group. Available at https://investors.natwestgroup.com/share-data/equity-ownership-statistics.aspx (2020)
	
	\bibitem{petroneLatora_dynamic_2018}
	Petrone,~D., Latora,~V. A dynamic approach merging network theory and credit risk techniques to assess systemic risk in financial networks. 
	\emph{Sci Rep} \textbf{8}, 5561 (2018).
	
	\bibitem{altman_default_2004}
	Altman,~E., Resti,~A. \& Sironi,~A. Default recovery rates in
	credit risk modelling: a review of the literature and empirical evidence.
	\emph{Econ. Notes} \textbf{33}, 183--208  (2004).
	
	\bibitem{CreditSuisse_2019}
	 Credit~Suisse, Available at https://www.credit-suisse.com/media/assets/corporate/docs/about-us/investor-relations/financial-disclosures/financial-reports/csg-ar-2019-en.pdf\#page=147 (2020)
	 
	\bibitem{Blix_2016}
	Blix Grimaldi,~M., Hofmeister,~J., Schich,~S., Snethlage,~D.
	Estimating the size and incidence of bank resolution costs for selected banks in OECD countries.
	\emph{OECD Journal: Financial Market Trends} \textbf{1}, (2016) 
	
	\bibitem{Cariboni_reducing_2016}
	Cariboni,~J., Fontana,~A., Langedijk,~S., Maccaferri,~S., Pagano,~A., Giudici~M. , Rancan,~M., Schich,~S. Reducing and sharing the burden of bank failures.
	\emph{OECD Journal: Financial Market Trends} \textbf{2}, (2016) 
	
	\bibitem{Caccioli_Network_2018} 
	Caccioli, F., Barucca, P. \&  Kobayashi, T. 
	Network models of financial systemic risk: a review. 
	\emph{J Comput Soc Sc} \textbf{1}, 81–114 (2018).
	
    \bibitem{gai_contagion_2010}
	Gai,~P. \& Kapadia,~S. Contagion in financial networks. \emph{Proc. Royal Soc. A}
	\textbf{466}, 2401--2423 (2010).
	
	\bibitem{haldane_managing_2014}
	 Haldane,~A.~G. Managing global finance as a system. Available at https://www.bankofengland.co.uk/speech/2014/managing-global-finance-as-a-system (2014).	
	
	\bibitem{boccaletti_complex_2006}
	Boccaletti,~S., Latora,~V., Moreno,~Y., Chavez,~M. \& Hwang,~D. Complex
	networks: Structure and dynamics. \emph{Phys. Rep.} \textbf{424},  175--308 (2006).
	
    \bibitem{wagalath_running_2011}
	Cont,~R. \& Wagalath,~L. Running for the exit: distressed selling and endogenous correlation in financial markets. \emph{Math. Finance} \textbf{23}, 718--741 (2013).
	
	\bibitem{battiston_debtrank_2012}
	Battiston, S., Puliga, M., Kaushik, R., Tasca, P. \& Caldarelli, G. {DebtRank}: too central to fail? financial networks, the {FED} and systemic risk. \emph{Sci. Rep.} \textbf{2}, 541 (2012).	
	
	\bibitem{okane_gaussian_2008}
	O'Kane,~D. The gaussian latent variable model in  \emph{Modelling
	Single-name and Multi-name Credit Derivatives} 241--259 (Wiley Finance, 2008).
	
	\bibitem{merton_pricing_1974}
	Merton,~R.~C. On the pricing of corporate debt: The risk structure of
	interest rates. \emph{J. Finance} \textbf{29}, 449--470 (1974).	
	
    \bibitem{lehar_measuring_2005}
	Lehar,~A. Measuring systemic risk: A risk management approach.
	\emph{J. Bank. Finance} \textbf{29}, 2577--2603 (2005).
	
	\bibitem{furfine_interbank_2003}
	Furfine,~C. Interbank exposures: Quantifying the risk of contagion.
	\emph{J. Money Credit Bank.} \textbf{35}, 111--128 (2003).
	
    \bibitem{huang_framework_2009}
	Huang,~X., Zhou,~H. \& Zhu,~H. A framework for assessing the
	systemic risk of major financial institutions. \emph{J. Bank. Finance}
	\textbf{33}, 2036--2049 (2009).	
	
	\bibitem{upper_simulation_2011}
	Upper,~C. Simulation methods to assess the danger of contagion in
	interbank markets. \emph{J. Financial Stab.} \textbf{7}, 111--125 (2011).	
	
    \bibitem{upper_estimating_2004}
	Upper,~C. \& Worms,~A. Estimating bilateral exposures in the
	german interbank market: Is there a danger of contagion? \emph{Eur. Econ. Rev.} \textbf{48}, 827--849 (2004).	

	\bibitem{sutton_reinforcement_2018}
	Sutton,~R. S. \& Barto,~A. G. Reinforcement Learning: An Introduction. (MIT Press, 2018).
	
	\bibitem{Gordon_approximate_1999}
	Gordon,~G. \& Tom Michael Mitchell. Approximate solutions to markov decision processes. (1999).
	
	\bibitem{ohalloran_artificial_2019} 
	O'Halloran,~S. \& Nowaczyk,~N. An Artificial Intelligence Approach to Regulating Systemic Risk. \emph{Front. Artif. Intell.}\textbf{2}, 1--14 (2019).
	
    \bibitem{anand_filling_2015}
	Anand,~K., Craig,~B. \& Von~Peter,~G. Filling in the blanks:
	Network structure and interbank contagion. \emph{Quant. Finance} \textbf{15}
	 , 625--636 (2015).\
	
	\bibitem{kou_machine_2019}
	Kou,~G., Chao,~X., Peng,~Y., Alsaadi,~F. \& Herrera-Viedma,~E. Machine Learning Methods for Systemic Risk Analysis In Financial Sectors. \emph{Technol. Econ. Dev. Econ.} \textbf{25}, 1--27 (2019).
	
	\bibitem{bellman_markovian_1957}
	Bellman,~R. E.  A Markovian decision process. \emph{J. Math. Mech.} \textbf{6(5)}, 679--684 (1957).
	
	\bibitem{bellman_dynamic_1957}
	Bellman,~R. E. Dynamic Programming. ( Princeton University Press, 1957)
	
	
	
	
	
	
	
	
	
	
	
	
	
	
	
	
	
	

	
	
	
	
	
	
	

	
	
	

	
	
	
	

	
	
	
	

	
	
	
	

	
	

	
	
	
	
	
	

	

	
	
	
	
	
\end{thebibliography}

\end{document}